\documentclass[prb,twocolumn]{revtex4}

\usepackage{graphicx}

\begin{document}

\title{A note on Dolby and Gull on radar time and the twin ``paradox''}
\author{Antony Eagle}
\email{antony.eagle@philosophy.ox.ac.uk}
\affiliation{Exeter College, Oxford, OX1 3DP, United Kingdom}

\begin{abstract}
Recently a suggestion has been made that standard textbook
representations of hypersurfaces of simultaneity for the travelling
twin in the twin ``paradox'' are incorrect. This suggestion is false:
the standard textbooks are in agreement with a proper understanding of
the relativity of simultaneity.
    \end{abstract}
    
    \maketitle

    \section{Introduction}
    
    It is common wisdom that there is nothing paradoxical about the
    twin ``paradox'', and that appropriate attention to the
    unambiguous differences between the proper times of the travelling
    twin (whom we shall call ``Barbara'') and the stay-at-home twin
    (``Alex'') resolves any air of paradox. Indeed, the reader may
    well feel exasperated that any work is still being done on this
    fallacious argument.
    
    However, in a recent article in this journal,\cite{dgr} Dolby and
    Gull argue that the standard resolutions of the twin ``paradox''
    incorrectly answer the question of how the travelling twin should
    assign times of occurrence to distant events; that is, how Barbara
    should represent her hypersurfaces of simultaneity. They claim
    that proper application of ``radar time'' to the accelerated twin
    allows us to sort out an unambiguous time of happening to each
    event, from Barbara's perspective, and they go on to do just that.
 
 I have no complaint with their mathematics (though see
 \S\ref{s:convent}). My query, rather, is with the point of their
 exercise. For correct attention to both the concept of simultaneity
 and to what it means for an observer to assign a ``time of
 happening'' to an event shows that there was no problem with the
 standard textbook resolutions in the first place---though perhaps it
 is true that the textbook authors were not aware of the facts that
 render their work unproblematic.
 
 \section{Dolby and Gull's worry}
 
\begin{figure}
    \centering
    \includegraphics{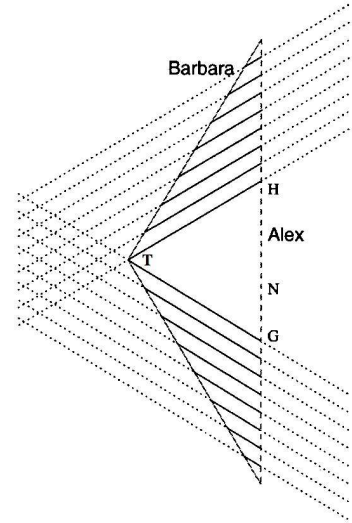}
    \caption{A typical ``textbook illustration'' of the hypersurfaces 
    of simultaneity of the travelling twin Barbara in the twin paradox.
    (Adapted from Dolby and Gull's Figure 1.)
  }
    \label{fig:1}
\end{figure}

Standard textbook resolutions of the twin paradox (in the
instantaneous turnaround case) claim that the hypersurfaces of
simultaneity for Barbara are as in Fig.~\ref{fig:1}.

\begin{figure}
     \centering
     \includegraphics{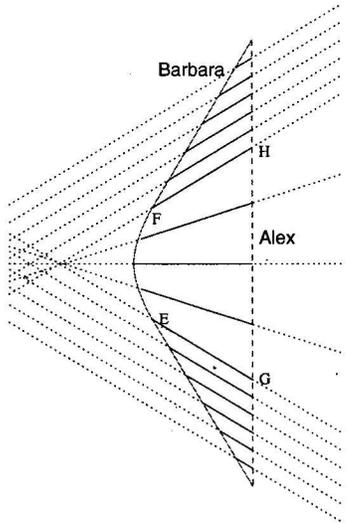}
     \caption{Another typical ``textbook illustration'', in which
     Barbara's hypersurfaces of simultaneity ``sweep around'' from
     $EG$ to $FH$ during the period of turnaround (acceleration).
     (Adapted from Dolby and Gull's Figure 2.)}
     \label{fig:2}
     \end{figure}

 Regarding this figure, Dolby and Gull point out two things. Firstly
 they note that when Barbara instantaneously turns around, the points
 $G$ and $H$ are regarded as simultaneous by Barbara. They acknowledge
 that this problem is dissolved if we make the situation more
 physically realistic, and adopt a turnaround of extended duration, as
 in Fig.~\ref{fig:2}. (We do well to note that this move still leaves
 untouched the conceptual problem in the original, instantaneous
 case.) Secondly, they note that moving to an extended turnaround
 ``cannot resolve the more serious problem\ldots which occurs to
 Barbara's left. Here her hypersurfaces of simultaneity are
 overlapping, and she assigns three times to every event!''\cite{dgr}
 They make no further comment, presuming that the exclamation mark is
 sufficient to convey their attitude to the standard treatments. That
 attitude, I assume, is that the very fact of assigning three times to
 every event is an absurdity---and the standard textbook resolutions
 must be wrong since they entail this absurd claim that there can be
 overlapping hypersurfaces of simultaneity.
 
\section{Defusing the worry}\label{def}

Is the claim that more than one time of happening can be assigned to a
single event such a patent absurdity as Dolby and Gull seem to regard
it? I will now give a {\em reinterpretation} of Fig.~\ref{fig:1},
which will show that it is far from an absurdity.

Let us call the point of turnaround $T$. Consider an observer,
$O_{1}$, who travels inertially until $T$ along the same path as
Barbara, and whose worldline then sadly terminates. Consider another
observer, $O_{2}$, who springs into existence at $T$, and shadows
Barbara thereafter.

I take it that it is obvious that the hypersurfaces of simultaneity
for $O_{1}$ and $O_{2}$ are as in Fig.~\ref{fig:1}, and that those
hypersurfaces overlap without problem or absurdity. Indeed, it is a
central fact about the relativistic theories that observers who are
moving differently are in different frames of reference, and hence
their partitions of events into simultaneity equivalence classes can
differ quite radically. So it cannot be the mere fact that the
hypersurfaces of simultaneity overlap that worries Dolby and Gull. It
must rather be that it is possible that {\em one observer}, who
happens to move differently at different times, can assign the same
event to different equivalence classes under simultaneity, again at
different times.

But, again, why should this be a problem? An absurd situation would
be: if at one and the same time, an observer assigned the same event
to two different hypersurfaces of simultaneity, since that would (at
the least) involve the absurdity that the observer was in two distinct
frames of reference at one and the same time, among other problems.
But there is no absurdity in the claim that, at different times,
whilst moving differently and hence in different frames, Barbara
assigns a single event to two different simultaneity classes. Hence,
it seems, there was no need to tinker with the original ``textbook''
presentations of her hypersurfaces of simultaneity: they are perfectly
adequate to capture Barbara's judgements of simultaneity at every time
at which she makes such a judgement.

Perhaps Dolby and Gull are worried that a single, persisting observer
cannot (without considerable mental problems) assign one and the same
event to different simultaneity classes, even if the observer never
does that at a single time. This seems an empty worry, for those
differing assignments are precisely what we should expect in a
relativistic scenario. Maybe Barbara is a determinedly
pre-relativistic individual, in which case she may have some
conceptual problems with the different assignments, perhaps because
she still operates with a hidden Newtonian assumption that there is
one universal time. But that cannot be at the root of Dolby and Gull's
worry, as they make no such assumption---see, however,
\S\ref{s:further}.

Dolby and Gull go on to say the following:\cite{dgr} \begin{quote}
[I]f Barbara's hypersurfaces of simultaneity at a certain time depend
so sensitively on her instantaneous velocity as these diagrams
suggest, then she would be forced to conclude that the distant planets
swept backwards and forwards in time whenever she went
dancing!\end{quote} Perhaps if we can decipher this remark, we will
get at the root of their worries.

As far as I can tell, their worry here is that, as Barbara's
instantaneous velocity changes from moment to moment, she will be
forced to conclude that some events that are in her current subjective
future (i.e.\ that lie within the future light cone of some event on
her current hypersurface of simultaneity) were, at some point on her
past worldline, judged to be in the past (i.e.\ lying within the past
light cone of some event on her past (then-current) hypersurface of
simultaneity). Of course, this is no absurdity: it has long been clear
that the pre-theoretical concepts of ``past'' and ``future'' do not
mesh perfectly with their relativized versions. Yet I cannot see
anything more to Dolby and Gull's worry, other that it is motivated by
pre-theoretical intuitions about distant assignments of pastness and
futurity to events, intuitions that should by now be seen as very
doubtful in a relativistic universe.

One residual worry remains: What happens at $T$? In Fig.~\ref{fig:1},
it does seem that, at $T$, Barbara has two conflicting simultaneity
assignments (this is related to Dolby and Gull's first worry). The
resolution is simple: we need to assign her at most one instantaneous
frame of reference. The obvious one to choose is that at the instant
$T$, she is counted as at rest; her frame of reference then yields a
hypersurface of simultaneity that is orthogonal to Alex's worldline,
running horizontally across the page. This assignment has the virtue
that it shows Fig.~\ref{fig:1} to be the limit of Fig.~\ref{fig:2}, as
the period of acceleration decreases in extension. Perhaps, though, we
might think that at this instantaneous point, there is no sense to be
made of the observer's frame of reference, and hence perhaps we assign
no hypersurface of simultaneity. This latter option has problems of
its own, of course; but in principle, either choice serves to resolve
the residual worry.

 \section{The Conventionality of Simultaneity} \label{s:convent}

 On the basis of the above considerations, I see no force to the
 motivating remarks that Dolby and Gull provide, and hence I query
 whether their mathematical work needed to be performed.
 
 Setting that issue aside, however, some interesting details emerge
 when one considers their positive proposal. They begin by defining
 the \emph{radar time} of an event $e$ basically as follows: let
 $t_{1}$ be the (proper) time at which the observer sends a signal to
 $e$, and let $t_{2}$ be the (proper) time at which the observer
 receives a return signal from $e$. The radar time $\tau(e)$ of $e$ is
 defined in equation \ref{eq:standard}.
 \begin{equation} \label{eq:standard} \tau(e) = t_{1} +
 \frac{1}{2}(t_{2} - t_{1}).\end{equation}
A hypersurface of simultaneity $\sigma_{e}$ is set of events with the
same radar time ($\sigma_{e} = \{x: \tau(x) = \tau(e)\}$); it is
obviously an equivalence class. This same relation of simultaneity is,
as Dolby and Gull are well aware, Einstein's standard convention for
simultaneity.\cite{ein}

Quite a large body of work has sprung up concerning the status of this
definition of simultaneity.\cite{mal,drt,ssd} Though they make passing
reference to some of this work,\cite{drt} Dolby and Gull do not engage
more thoroughly with it. If they had, they would have noticed that
while their radar time definition of simultaneity is an acceptable
definition, it is by no means the only available option.

Debs and Redhead\cite{drt} maintain that any definition of radar time
is acceptable if it is compatible with the following:
\begin{equation}\label{eq:convsim}
\tau_{\epsilon}(e) = t_{1} + \epsilon (t_{2} - t_{1}) \quad (0 <
\epsilon < 1).\end{equation} They take it that any particular choice
of value of $\epsilon$ in equation \ref{eq:convsim}
is {\em conventional}: that is, not fixed by the physical facts,
but rather by our conventional decision to use the term
``simultaneous'' to pick out an equivalence class under
$\tau_{\epsilon}$. No contradiction with any physical fact is possible
for any of these relations defined by different $\epsilon$ values,
because only proper time has ``objective status in special
relativity.''\cite{fri} If Debs and Redhead are right, then no special
significance will attach to assignments of distant simultaneity at
all: they are arbitrary and hence without physical importance. If that
is true, then the purported conflict over assignments which so
exercised Dolby and Gull is of even smaller significance than I made
out above.

Of course, others have rushed to defend the Einsteinian convention,
proving its unique adequacy if we set certain conditions on a
plausible candidate for simultaneity.\cite{mal,rob} But those further
conditions have been disputed too.  Though I have no wish to defend it
here, it seems that conventionalism about simultaneity remains an open
possibility.\cite{drt,ssd} If that possibility turns out to be true,
then Dolby and Gull have done some excellent work defining a potential
candidate simultaneity relation, but one that loses its importance
once we see that any candidate relation, within wide bounds, will do.
In particular, the textbook presentations of the hypersurfaces of
simultaneity are perfectly good candidates for us to adopt as our
convention.

\section{Further External Constraints} \label{s:further}

However, it seems that Dolby and Gull's motivation for proposing a
unique foliation for any given observer does not rest on
considerations intrinsic to special relativity.\cite{gco,dgs} Rather,
they have a particular \emph{application} in mind, for which they
regard unique foliation as crucial. The particular application is to
fermionic particle creation in relativistic quantum field theory.
Obviously if this application requires unique foliation, then these
remarks about conventionality of simultaneity will be misapplied,
since we are now in effect imposing \emph{additional} physical
constraints that suffice for uniqueness. Moreover, their particular
proposal requires \emph{consistency} of foliation for a given
observer: so the textbook overlapping hypersurfaces at different times
will also be incorrect.

One small concern about this approach is that we have strictly gone
beyond the conceptual content of special relativity, and therefore
that it is inappropriate to represent the Einsteinian convention as
the only one possible for special relativity. Rather, the Einstein
proposal is the only one that it is \emph{pragmatically appropriate}
to use when applying special relativity, given the further constraints
imposed by our beliefs concerning what the actual world is like. But
strictly speaking simultaneity may yet be conventional in standard
special relativity without external constraints.

A bigger worry, however, is with the particular application they have
in mind. The crucial point for Dolby and Gull is that, if we cannot
establish a unique foliation for a given observer, certain features of
the \emph{particle distribution} for that observer will change. They
argue that it is extremely odd indeed that a mere change in velocity
(or worse, a merely conventional choice of simultaneity relation)
could result in the appearance or disappearance of particles.

Unfortunately, it is now quite clear that there is no relativistic
quantum theory of localizable particles.\cite{hcn,mald} The most
plausible response to this result, and indeed the standard response,
is to argue that since different observers see different particle
distributions, we should be \emph{anti-realists} (or
conventionalists!) about particle distributions. If this standard
plausible response were accepted, that would undercut Dolby and Gull's
motivation for wanting a unique foliation of the spacetime to ensure
reliable and constant particle distributions.

In sum, if we are irrealists about particle distributions, we have no
motivation to demand unique foliations for given observers---and we
have every reason to be irrealist about particle distributions, given
the impossibility of giving an adequate characterization of local
particles within relativistic quantum field theory, and moreover given
that ``RQFT {\em itself} shows how the `illusion' of localizable
particles can arise, and how talk about localizable particles can be a
useful fiction.''\cite{hcn} Insofar as we are prepared to deploy a
particle number observable, we do so from the perspective of an
instantaneous rest frame of an observer, and the foliation consequent
upon that choice of rest frame, without demanding that this foliation be
constant for a given observer no matter what their motion.

\section{Conclusion}

Dolby and Gull have given an elegant and precise example of how to
apply the standard Einsteinian convention on assignments of distant
simultaneity in the twin paradox case. It very usefully illustrates
that the standard convention can be applied alike to accelerating and
inertial observers, contrary to some of the received wisdom in the
field.\cite{bon}

However, their work was not strictly necessary. Firstly because there
was no intuitive paradox or problem with the standard textbook
presentations, as I showed in \S\ref{def}. Secondly, many people
regard the definition of simultaneity as a conventional matter in any
case, as discussed in \S\ref{s:convent}. Hence no acceptable choice
for this convention can be criticized as mistaken on physical grounds,
but only on grounds of usefulness for our purposes. Thirdly, even if
we undermine the conventionality response by adducing some further
actual physical fact, that wouldn't show the conceptual commitments of
special relativity to be any different---moreover, I suspect that any
such additional physical fact will not in actuality conflict with the
standard representations of hypersurfaces of simultaneity
(\S\ref{s:further}). Given these observations, I suggest, the standard
textbook suggestion concerning what should be regarded as the
hypersurfaces of simultaneity is at least as successful as Dolby and
Gull's more complicated alternative proposal.

\begin{acknowledgments}
    Thanks to Hans Halvorson, Dave Baker, Steve Gull, and two anonymous reviewers for helpful
    comments, and to Luke Elson for the initial discussion which prompted my 
    interest in this material. 
   \end{acknowledgments}

\end{document}